\documentclass[11pt]{article}

\usepackage{hyperref}
\hypersetup{
    colorlinks=true,
    linkcolor=blue,
    filecolor=magenta,      
    urlcolor=blue,
}

\usepackage{color}
\usepackage{latexsym}
\usepackage{epsfig,graphics}
\usepackage{amsmath,amssymb,amsthm}
\usepackage{bm}

\DeclareMathAlphabet\mathbfcal{OMS}{cmsy}{b}{n}

\usepackage{upref}
\usepackage{setspace}
 \newcommand{\beqn}{\begin{eqnarray}}
 \newcommand{\eeqn}{\end{eqnarray}}
 \newcommand{\be}{\begin{equation}}
 \newcommand{\ee}{\end{equation}}
 \newcommand{\ba}{\begin{array}}
 \newcommand{\ea}{\end{array}}

 \newcommand{\pa}{\partial}
 
  \newcommand{\ci}{\cite}
 \newcommand{\ds}{\displaystyle}
 \newcommand{\la}{\label}

 \newcommand{\blangle}{{\big\langle}}
  \newcommand{\brangle}{{\big\rangle}}
 
 \newcommand{\fr}{\frac}
 
 \newcommand{\we}{{\rm \wedge}}

\newcommand{\ov}{\overline}

\newcommand{\cF}{{\cal F}}

\newcommand{\bA}{{\bf A}}

\newcommand{\cE}{{\cal E}}

\newcommand{\bH}{{\bf H}}

\newcommand{\cJ}{{\cal J}}

\newcommand{\cL}{{\cal L}}

\newcommand{\bx}{{\bf x}}

\newcommand{\cP}{{\cal P}}

\newcommand{\cV}{{\mathbb{V}}}

\newcommand{\bY}{{\bf Y}}
\newcommand{\cY}{{\mathbb{Y}}}

\newcommand{\ve}{\varepsilon}

\newcommand{\De}{\Delta}

\newcommand{\dv}{{\rm div\5}}
\newcommand{\curl}{{\rm curl\5}}

\newcommand{\al}{\alpha}

\newcommand{\om}{\omega}

\newcommand{\na}{\nabla}
\newcommand{\bPi}{{\bm\Pi}}

\newcommand{\5}{{\hspace{0.5mm}}}

\newcommand{\R}{\mathbb{R}}

\newtheorem{theorem}{Theorem}[section]

\renewcommand{\thetheorem}{\arabic{section}.\arabic{theorem}}
\newtheorem{defin}[theorem]{Definition}

\newtheorem{lemma}[theorem]{Lemma}
\newtheorem{remark}[theorem]{Remark}

\newtheorem{pro}[theorem]{Proposition}
\newcommand{\bp}{\begin{pro}}
\newcommand{\ep}{\end{pro}}

\newcommand{\bt}{\begin{theorem}}
\newcommand{\et}{\end{theorem}}

\newcommand{\bl}{\begin{lemma}}
\newcommand{\el}{\end{lemma}}

\newcommand{\const}{\mathop{\rm const}\nolimits}

\newcommand{\bce}{\begin{center}}
\newcommand{\ece}{\end{center}}

\newcommand{\bpr}{\begin{proof}}
\newcommand{\epr}{\end{proof}}

\newcommand{\br}{\begin{remark}}
\newcommand{\er}{\end{remark}}

\newcommand{\bd}{\begin{defin}}
\newcommand{\ed}{\end{defin}}

\begin{document}

\bce
{\Large Momentum map for the Maxwell--Lorentz equations  
 \medskip
 
with spinning particle}
 \bigskip \medskip

 {\large A.I. Komech}
 \\
{\it
Faculty of Mathematics,  Vienna University\\
}
 alexander.komech@gmail.com
\smallskip

 {\large E.A. Kopylova}\footnote{ 
 Supported partially by Austrian Science Fund (FWF) P34177
 }
 \\
{\it
Faculty of Mathematics,   Vienna University
    }
  \\
  elena.kopylova@univie.ac.at
\ece

\begin{abstract}

We develop the theory of momentum map for the Maxwell--Lorentz equations with spinning extended charged particle. This theory is indispensable for the study of long-time behaviour and radiation of the solitons of this system. The development relies on the 
Hamilton--Poisson structure of the system. As an example, we apply the theory to the  rotation group of symmetry, calculating the expression for the conserved angular momentum. We check the coincidence of this expression  with known classical invariant.

\end{abstract}

  \bigskip

MSC 35Q61;37K06; 37K40;53D17; 70S05;70G65;70G45.

\newpage
\tableofcontents

\section{Introduction}

Our goal is a development  of the methods of Geometric Mechanics \ci{A1966,AK1998,H2009,K1996,MMOPR2007,MR2002} in the 
context of  the Maxwell--Lorentz equations
with spinning extended charged  particle.
The development relies on 
 the Hamilton--Poisson  representation
of the equations established in \ci{KK2023hp}
with application of the Lie--Poincar\'e calculus 
\ci{A1966,H2009}.

 The system is invariant with respect  
to the translation and rotation groups, that
allows one to apply the momentum map theory 
developed in the Geometric Mechanics.

We choose the units where the speed
of light is $c=1$.
Then the  Maxwell--Lorentz equations with a nonrelativistic spinning particle read
(see \ci{K1999,N1964,S2004})
\begin{equation} \la{mls3}
\left\{\begin{array}{rcl}
\dot E(x,t)\!\!&\!\!=\!\!&\!\!\curl B(x,t)-w(x,t)\rho(x-q(t)),\qquad \dot B(x,t)= - \curl E(x,t)
\\\\
\dv E(x,t)\!\!&\!\!=\!\!&\!\!\rho (x-q(t)),\qquad\qquad\qquad\quad \quad \quad\,\,
\dv B(x,t)=0
\\\\
m\ddot q(t)\!\!&\!\!=\!\!&\!\!\blangle E(x,t)+w(x,t)\we B(x,t),\rho(x-q(t)) \brangle
\\\\
 I\dot \om(t)\!\!&\!\!=\!\!&\blangle (x-q(t))\we
\big[E(x,t)+w(x,t)\we B(x,t)\big],
\rho(x-q(t))\brangle
\end{array}\right|, \qquad
\end{equation}
where $w(x,t)$ is the velocity field
\be\la{w}
w(x,t):=\dot q(t)+\om(t)\we(x\!-\!q(t)).
\ee
We denote by
$m$  the mass of the particle and by $I$ its moment of inertia,
$\rho(x-q)$ is the charge distribution of the extended
particle centered at a point $q\in\R^3$, and 
the brackets $\blangle,\brangle$ denote the inner product in the  Hilbert space
$L^2:=L^2(\R^3)\otimes\R^3$. Further, $\om(t)$ is the angular velocity 
of the particle rotation: every fixed point $x(0)\in\R^3$ of the extended particle
moves along the trajectory $x(t)=q(t)+R(t)(x(0)-q(0))$, where $R(t)\in SO(3)$, and the
velocity of the point is 
\beqn\la{tra}
\dot x(t)&=&\dot q(t)+
\dot R(t)(x(0)-q(0))=\dot q(t)+\dot R(t)R^{-1}(t) (x(t)-q(t))
\nonumber\\
\nonumber\\
&=&\dot q(t)+\om(t)\we (x(t)-q(t)),\qquad \om(t)\we=\dot R(t)R^{-1}(t)
\eeqn
($\om(t)$ is called the {\it angular velocity in the space}, see \ci{A1989}).

The system (\ref{mls3}) is invariant under the action of
the group of translations and rotations of the space $\R^3$
if the charge density $\rho(\bx)$ is rotation invariant.
Accordingly, the system  admits solitons
 constructed in \ci{S2004},
that move with constant speed and rotate with constant angular velocity.
\smallskip
 
 The main result of the present paper is the
 development of the theory of momentum map 
 in the framework of  the system (\ref{mls3})
 and the groups of  rotations
  following  methods of Geometric Mechanics
\ci{A1966,AK1998,H2009,MMOPR2007,MR2002}.
The development 
  became possible due to  explicit calculation
in \ci{KK2023hp} 
of the skew-symmetric {\it structural operator},
which is the ``integral kernel" of
the Poisson bracket.  The calculation relies on the Lie--Poincar\'e
calculus  \ci{L1890,P1901}.

The theory of momentum map is of crucial 
importance in applications to nonlinear 
Hamiltonian  PDEs since it gives a systematic approach to the construction of invariants. 
The invariants play
 a key role in the process of  reduction 
 for
 the analysis of stability questions \ci{GSS,KK2023r}. 
 This fact was the main inspiration for this article.

\smallskip  
The system (\ref{mls3}) without rotation (the first three lines with $w(t)=\dot q(t)$)
 plays a crucial role in a rigorous analysis of radiation by moving solitons, see
  \ci{BG1993, IKM2004,
  KK2020,KK2022,KKS1999,
  KS1998,KS2000,KSK1997,
 K2010,KS2000ad, S2004}. We expect that the results
 of present article will help to extend this analysis 
 to the 
  moving and rotating  solitons
 for the
 system (\ref{mls3}).

\smallskip

Let us comment on related results.

  The coordinate-free proof of the Hamilton least action principle
was given in
\ci{IKS2015} on the basis of   the 
technique of \ci{AKN1997,P1901}:
 the  Poincar\'e equations, and 
 expansions over right-invariant 
  vector fields on the Lie group $SO(3)$. 
 This technique was developed 
in \ci{IKS2017}, where the general theory of invariants was constructed
for the Poincar\'e equations on manifolds and applied to
the construction of  invariants for the system (\ref{mls3}). 
The coordinate-free proof of the conservation laws for the system (\ref{mls3})
was given by Kiessling \ci{K1999}.

\setcounter{equation}{0}
\section{The Lagrangian structure}
In this section we recall the
 Lagrangian structure and well-posedness
for  system (\ref{mls3}). This means that the equations can be obtained from the Hamilton 
least action principle.
Denote the Sobolev spaces $H^s=H^s(\R^3)\otimes\R^3$ with $s\in\R$,
and  $\dot H^1=\dot H^1(\R^3)\otimes\R^3$.
All the 
derivatives are understood in the sense of distributions.
 We assume that the charge density $\rho(x)$ is smooth and  spherically-invariant, i.e.,
\be\la{rosym}
\rho\in C_0^\infty(\R^2),\qquad\rho(x)=\rho_1(|x|), \qquad 
\rho(x)=0\,\,\,{\rm for}\,\,\,|x|\ge R_\rho.
\ee

\subsection{The Maxwell potentials}
In the Maxwell potentials $A(x,t)=(A_1(x,t),A_2(x,t),A_3(x,t))$ and $\Phi(x,t)$, we have
(see \ci{Jackson})
\be\la{AA}
B(x,t)=\curl A(x,t),\qquad E(x,t)=-\dot A(x,t)-\na \Phi(x,t).
\ee
We choose the Coulomb gauge
\be\la{Cg}
\dv A(x,t)=0.
\ee
\bd The Hilbert space 
$\cF^0:=\{A\in L^2: \dv A(x)\equiv 0\}$, and $\dot\cF^1$
is the completion of $\cF^0\cap [C_0^\infty (\R^3)\otimes\R^3]$
w.r.t. the norm 
\be\la{cF}
\Vert\psi\Vert_{\dot\cF^1}^2=\int |\na\psi(x)|^2dx<\infty.
\ee
\ed
 Now  the system  (\ref{mls3}) becomes
(see \ci{KK2023hp})
\be\la{2ml2}
\left\{\ba{rcl}
\ddot A(x,t)\!\!&\!\!=\!\!&\!\!\De A(xmt)+\cP \big([\dot q(t)+\om\we (x-q(t))] \rho(x-q(t)) \big)
\\\\
m\ddot q(t)\!\!&\!\!=\!\!&\!\!\blangle-\dot A(x,t)+[\dot q(t)+\om\we (x-q(t))]\we\curl A(x,t) ,\rho(x\!-\!q(t)) \brangle
\\\\
 I\dot \om(t)\!\!\!&\!\!=\!\!&\!\!\blangle (x\!-\!q(t))\we 
 \Big[-\dot A(x,t)+ [\dot q(t)+\om(t)\we (x-q(t))]\we\curl A(x,t)\Big],\rho(x\!-\!q(t))\brangle
\ea\right|.
\ee
where
the brackets $\blangle,\brangle$ denote the inner product in the real Hilbert space
$L^2$.

\br\rm

 For the system without rotation (the first two equations with $\om=0$) the Lagrangian structure was known more than
 one hundred years \ci{LL1975}. For the system with rotation,
 the Lagrangian structure 
 was established for the first time in \ci{IKS2015} using the 
 Poincar\'e equations on the rotation group $SO(3)$, \ci{AKN1997,P1901}.

\er

\subsection{The well-posedness}\la{WA}
Here we recall
the well-posedness 
for the system (\ref{2ml2}).
Denote the Hilbert spaces
\be\la{cYW}
\cY:=\dot\cF^1\oplus\cF^0\oplus \R^3\oplus \R^3\oplus \R^3,
\qquad
\cV=L^2\oplus H^{-1}\oplus\R^3\oplus \R^3\oplus \R^3.
\ee
\bp\la{pwpA}
{\rm i)} For any initial state $Y(0)=(A(x,0),\dot A(x,0),q(0),p(0),\om(0))\in\cY$,  the system {\rm (\ref{2ml2})} admits a unique
solution 
\be\la{soluY}
Y(t)=(A(x,t),\dot A(x,t),q(t),p(t),\om(t))\in C(\R,\cY)\cap C^1(\R,\cV).
\ee
{\rm ii)} The map 
$
W(t):Y(0)\mapsto Y(t)
$
 is continuous in $\cY$ for every $t\in\R$.
 \\
{\rm iii)} Let 
\be\la{AP0}
\left\{\ba{l}
A(x,0)\in C^3(\R^3)\!\otimes\!\R^3,\quad \dot A(x,0)\in C^2(\R^3)\!\otimes\!\R^3
\\\\
A(x,0)=0,\,\,\,\dot A(x,0)+\na\Phi(x,0)=0,\qquad |x|>R
\ea\right|
\ee
for some $R>0$. Then
\be\la{AP1}
\left\{\ba{l}
A(x,t)\in C^2(\R^3\times\R)\!\otimes\!\R^3,
\\\\
  |\pa_x^\al A(x,t)|+|\pa_x^\al \dot A(x,t)|\le C_\al(1+|x|)^{-2-|\al|},\quad |x|>r(t)+1,\qquad \forall\al
\ea\right|,
\ee
where 
$r(t)=R+|q(0)|+\ov v|t|+R_\rho$ with
$\ov v>0$ defined by the relation $\cE=\fr12m\ov v^2$.
For such solutions,
the angular momentum is conserved:
\be\la{amcoA}
J=\blangle A(x,t)\we \Pi(x,t)\brangle-
\blangle ((x-q)\we \na)_*A(x,t),\Pi(x,t)\brangle+q(t)\we P+\pi(t)=\const,
\ee
where $\pi(t)=I\om(t)+\blangle (x-q(t))\we A(x,t),\rho(y)\brangle
$ and
 by definition, 
\be\la{na*}
\blangle ((x-q)\we \na)_*A(x,t),\Pi(x,t)\brangle_n=\blangle ((x-q)\we \na)_nA(x,t),\Pi(x,t)\brangle,
\qquad n=1,2,3.
\ee
\ep
The assertions i) and ii) have been proved in   \ci{KK2023hp}, while iii) will be proved
in the present paper.

\br 
{\rm
In Section \ref{D}, we check that the expression
for the angular momentum (\ref{amcoA}) 
coincide
with the corresponding classical invariant (\ref{amco}).
}
\er


\setcounter{equation}{0}
\section{The Hamilton--Poisson structure}
Let us denote $\Pi$, $p$, and $\pi$ are the conjugate momenta 
\be\la{pi}
\Pi:=\dot A,\qquad 
p:=m\dot q+\blangle A(q+y),\rho(y)\brangle,
\qquad\pi=I\om+\blangle y\we A(q+y),
\rho(y)\brangle.
\ee
In \ci{KK2023hp}, we have shown that the system  (\ref{2ml2})
admits a representation in the Hamiltonian form
\be\la{Ham2}
\left\{\ba{rcl}
\dot A&=&D_\Pi H,\qquad\dot \Pi=-D_A H
\\\\
\dot q&=&D_p H,\quad \,\, \dot p=-D_q H
\\\\
 \dot \pi&=& -\pi\we D_\pi H
\ea\right|,
\ee
where $H$ is the Hamiltonian
 \beqn\la{Ham}
\!\!&\!\!&\!\!H(A,\Pi,q,p,\pi)=
\fr12\blangle\Pi,\Pi\brangle+
\fr12\blangle\curl A,\curl A\brangle+ \fr12m\dot q^2+\fr12I\om^2
\nonumber\\
\nonumber\\
\!\!&\!\!=\!\!&\!\!
\fr12\int[\Pi^2+(\curl A)^2]dx+\fr1{2m} [p-\blangle A(x),\rho(x-q)\brangle]^2
+\fr1{2I}[\pi-\blangle (x-q)\we A(x),\rho(x-q)\brangle]^2,\qquad
\eeqn
The Hamiltonian $H$ is well defined
and Fr\'echet differentiable
on the Hilbert phase space  $\cY$ introduced in (\ref{cYW}).
Obviously,
the system (\ref{Ham2}) admits the representation
\be\la{Hf}
\dot Y=\cJ(Y) D H(Y),\qquad Y=(A,\Pi,q,p,\pi),
\ee
where  
$\cJ(Y)$ is a skew-symmetric {\it structural operator}:
\be\la{HsJ}
\cJ(Y)=\left(
\ba{ccccc}
0&1&0&0&0\\
-1&0&0&0&0\\
0&0&0&1&0\\
0&0&-1&0&0\\
0&0&0&0&-\pi\we
\ea
\right),\qquad Y=(A,\Pi,q,p,\pi).
\ee

\br
\rm
The Hamiltonian form (\ref{Ham2}), (\ref{Hf})  of the system (\ref{2ml2}) 
is obtained in
\ci{KK2023hp}
 by the Lie--Poincar\'e calculus  \ci{L1890, P1901}
from the Hamilton least action principle established
in \ci{IKS2015}.

\er

\setcounter{equation}{0}
\section{Momentum map for the rotation group $SO(3)$}
 
The conservation of energy, momentum
and angular momentum 
 for the system  (\ref{mls3}) (and  hence, for (\ref{2ml2}) and (\ref{Ham2})) are known, see \ci{K1999, IKS2017} for the formal proofs. 
All the systems (\ref{mls3}), (\ref{2ml2}), and (\ref{Ham2})      
are invariant with respect to rotation group $SO(3)$.  
However, 
the angular momentum
cannot be constructed with the Noether theorem 
since the rotation group is not a symplectic manifold. 

In this section, we develop the theory of momentum map \ci{A1966,H2009,MMOPR2007,MR2002}
corresponding to the symmetry group $SO(3)$ and apply it  
for the calculation of the conserved angular momentum.
 The development relies on the 
Hamilton--Poisson structure (\ref{Ham2}).

In Appendix \ref{D} we check that the obtained expression (\ref{Jam}) 
coincides with known
invariant (\ref{amco}) (its conservation was proved in \ci{K1999}).

\subsection{Canonical transformation to the comoving frame}
To consider the rotation invariance of 
 the system (\ref{Ham2}), we transform the system  to the comoving frame. 
Denote the fields in the comoving frame
 \be\la{fcom}
 \bA(y,t):=A(q+y,t), \qquad \bPi(y,t):=\Pi(q+y,t).
 \ee
 We  are going to change the variables in the system (\ref{Hf})
 via the map
 \be\la{T}
T: (A,q,\Pi,p,\pi)\mapsto (\bA,q,\bPi,P,\pi),
\ee
where $P$ is the conserved total momentum \ci{KK2023hp}:
\be\la{PP2}
P
=p-\langle \bPi, \na_* \bA\rangle,
\ee
Now we define
\be\la{Ham2d}
\bH(\bA,\bPi,q,P,\pi):=H(A,\Pi,q,p,\pi).
%
\ee
In \ci[Lemma 4.1]{KK2023q}, we have proved that $T$ is  the 
canonical transformation, so 
the system (\ref{Hf}) in the new variables is equivalent to
similar system with the Hamiltonian $\bH$:
\be\la{Hf2}
\dot \bY=J(\bY) D_\bY \bH(\bY),\qquad \bY=(\bA,\bPi,q,P,\pi),
\ee 
where $J(\bY)$  coincides with the right hand side of  (\ref{HsJ}).

\subsection{Action of the rotation group}

Using the momentum conservation (\ref{PP2}), we can rewrite Hamiltonian (\ref{Ham2d}) as
\beqn\la{Hamco}
&&\bH(\bA,\bPi,q,P,\pi)=\fr12\int[\bPi^2+(\curl \bA)^2]dx
\nonumber\\
\nonumber\\
&+&\fr1{2m} [
P+\langle \bPi, \na_* \bA\rangle
-\blangle \bA(y),\rho(y)\brangle]^2
+\fr1{2I}[\pi-\blangle y\we \bA(y),\rho(y)\brangle]^2\qquad\qquad
\eeqn
 For the system  (\ref{Hf2}), the action of the rotation group
 is defined as follows: for $R\in SO(3)$, 
\be\la{TR}
T(R)\bY:=(R\bA(R^{-1}y),R\bPi(R^{-1}y),
Rq,RP,
R\pi),\qquad \bY=(\bA,\bPi,q,P,\pi).
\ee
Due to  (\ref{rosym}),
the Hamiltonian (\ref{Hamco}) is invariant with respect to this action:
\be\la{TRH}
\bH(T(R)\bY)=\bH(\bY),\qquad R\in SO(3).
\ee
We check this invariance in Appendix \ref{aIH}.
The vector field of the deformation 
corresponding to the one-parametric subgroup of rotations $e^{s\hat\xi}$
is defined as
\be\la{TRf}
v_\xi(\bY):=\fr d{d\ve}\Big|_{s=0}T(e^{s\hat\xi}) \bY,\qquad 
\hat\xi:=\left(\ba{ccc}
0&-\xi_3&\xi_2\\
\xi_3&0&-\xi_1\\
-\xi_2&\xi_1&0
\ea
\right),\qquad \xi\in\R^3.
\ee
Differentiating, we obtain that
\be\la{TRf2}
v_\xi(\bY)=(\xi\we \bA-((\xi\we y)\!\cdot\!\na)\bA,
\xi\we \bPi-((\xi\we y)\!\cdot\!\na)\bPi,\xi\we q,\xi\we P,
\xi\we\pi).
\ee
\subsection{The momentum map}
The corresponding momentum map $J:\cY\to so^*(3)\simeq \R^3$ is defined by the equation
\be\la{momJ}
F_{J_\xi}(\bY)=v_\xi(\bY),\qquad \bY\in \cY,
\ee
where $J_\xi:=J\cdot\xi$ and 
\be\la{FJxi}
F_{J_\xi}(\bY)=\cJ(\bY)D J_\xi
\ee
is the Hamilton field corresponding to the Hamilton
 functional $J_\xi$.
  Using the structure
 operator (\ref{HsJ}) of the system (\ref{Hf2}), we obtain the Hamilton field
 \be\la{Haf}
 F_{J_\xi}(Y)=
 (D_\bPi J_\xi,-D_\bA J_\xi,
 D_P J_\xi,-D_q J_\xi,
 -\pi\we D_\pi J_\xi).
 \ee
Substituting these expressions and (\ref{TRf2}) into equation (\ref{momJ}),
we obtain the system of variational equations for the functional $J_\xi$:
\be\la{momJ2}
\left\{\ba{rclrcl}
D_\bPi J_\xi\!\!&\!\!=\!\!&\!\!\xi\we \bA-((\xi\we y)\!\cdot\!\na)\bA,&
-D_\bA J_\xi\!\!&\!\!=\!\!&\!\!\xi\we \bPi-((\xi\we y)\!\cdot\!\na)\bPi
\\\\
D_P J_\xi\!\!&\!\!=\!\!&\!\!\xi\we q,&
-D_q J_\xi\!\!&\!\!=\!\!&\!\!\xi\we P
\\\\
-\pi\we D_\pi J_\xi\!\!&\!\!=\!\!&\!\!\xi\we\pi
\ea\right|.
\ee
The solution is easy:
\be\la{Jxi}
J_\xi=
\blangle\xi\we \bA-((\xi\we y)\!\cdot\!\na)\bA,\bPi\brangle+(\xi\we q)\cdot P
+\xi\!\cdot\!\pi.
\ee
This implies the angular momentum (\ref{amcoA}):
\be\la{Jam}
J=\blangle \bA\we\bPi\brangle-
\blangle (y\we \na)_*\bA,\bPi\brangle+q\we P+\pi,
\ee
where the expression with $\na_*$ is defined similarly to (\ref{na*}):
in the components,
\be\la{Jamc}
J_n=\blangle \bA\we\bPi\brangle_n-\blangle (y\we\na)_n\bA,\bPi\brangle+(q\we P)_n+\pi_n,\qquad n=1,2,3.
\ee
\bp
The functional (\ref{Jam}) is conserved for solutions of the  system
(\ref{Hf2}) with initial data satisfying conditions (\ref{AP0}).

\ep
\bpr
The Hamiltonian
 (\ref{Hamco}) is invariant with respect to the action (\ref{TR}) of the rotation group.
 Hence, for any $\xi\in so(3)\simeq \R^3$, the Lie derivative in the direction
 of the vector field $v_\xi$ vanishes:
 \be\la{Hinv}
 \cL_{v_\xi}\bH=0.
 \ee
 Therefore, for any solution $\bY(t)$ of the  system
(\ref{Hf2}), using (\ref{FJxi}) and (\ref{momJ}), we obtain
\be\la{Hinv2}
\fr d{dt}J_\xi(\bY)=\blangle D J_\xi(\bY),\dot\bY \brangle=\blangle D J_\xi,\cJ(\bY)D\bH\brangle=-\blangle \cJ(\bY)D J_\xi,D\bH\brangle=
 -\cL_{v_\xi}\bH=0,
  \ee 
  where all 
  expressions are well-defined 
  and the identities hold
  by (\ref{AP1}). 
\epr
 
In the next section, we check that 
the angular momentum  (\ref{Jam}) 
coincides with known classical invariant
 (\ref{amco}).

\setcounter{equation}{0}
\section{Comparison with classical invariant}\la{D}
Here we check the coincidence of 
the expressions for the 
 angular 
momentum (\ref{amcoA}) 
with the corresponding classical invariant \ci{K1999}:
\beqn\la{eco}
J_c=\int x\we[E(x)\we B(x)]dx+I\om+q
\we m\dot q. \la{amco}
\eeqn
The energy
momentum coincide for all solutions $C(\R,\cY)$, while 
angular momentum coincide for solutions with initial data 
satisfying (\ref{AP0}). 
The coincidence of all invariants 
it suffices to prove for the case (\ref{AP0}), so all
integration by parts below are correct by  (\ref{AP1}).
We must to show that 
\be\la{JJc}
J_c= J.
\ee
Substituting (\ref{pi}), we rewrite  (\ref{amcoA})  as
\be\la{Jam22}
J=\blangle \bA\we\bPi\brangle-
\blangle (y\we \na)_*\bA,\bPi\brangle
+
q\we P+I\om+\blangle y\we \bA(y),\rho(y)\brangle.
\ee
Now
compare  (\ref{amco}) with classical form of the conserved momentum (\ref{PP2}):
\be\la{mco}
P_c=\ds\int E(x)\we B(x)dx+m\dot q.
\ee
Then
(\ref{amco}), can be  rewritten  as
\beqn\la{amco2}
J_c&=&q\we \int (E\we B)dx
+
\int (x-q)\we(E\we B)dx+I\om+q\we m\dot 
\nonumber\\
\nonumber\\
&=&
q\we (P-m\dot q)
+
\int (x-q)\we(E\we B)dx+I\om+q\we m\dot q
\nonumber\\
\nonumber\\
&=&
q\we P
+
\int (x-q)\we(E\we B)dx+I\om.
\eeqn
Hence,
it remains to check that 
\be\la{Jam3}
\int (x-q)\we(E(x)\we B(x))dx=
\blangle \bA\we\bPi\brangle-
\blangle (y\we \na)_*\bA,\bPi\brangle
+\blangle y\we \bA(y),\rho(y)\brangle.\qquad
\ee
Substituting (\ref{AA}), we obtain the equivalent identity
\be\la{Jam4}
-\int y\we[(\bPi(y)+\na \Phi(y))
\we \curl \bA(y)]dy
=
\blangle \bA\we\bPi\brangle-
\blangle (y\we \na)_*\bA,\bPi\brangle
+\blangle y\we \bA(y),\rho(y)\brangle.\qquad
\ee
Integrating by parts, and using that $\dv \bPi(y)=0$, we obtain
\be\la{Jam6}
\int y\we[\bPi(y)
\we \curl \bA(y)]dy
=- \blangle (y\we\na)_*\bPi,\bA\brangle+\blangle\bPi\we\bA\brangle.
\ee
It remains to prove that
\be\la{ek0}
-\int y\we\big([\na \Phi(y)]\we \curl \bA(y)\big)dy=\int y\we \bA(y)\rho(y)dy
\ee
The identity (\ref{ek0}) follows by a partial integration using that
 $\dv \bA(y)=0$ and $\De\Phi(y)=-\rho(y)$.
 Indeed,
 the formula bac-cab gives
\be\la{ek}
-\int y\we(\na\Phi \we \curl \bA) dy =\langle (y\cdot \na\Phi),\curl \bA\rangle-\langle \na\Phi, (y\cdot \curl \bA)\rangle
\ee
since $\Phi$ is real. 
Integrating by parts, we obtain
\be\la{ek1}
\langle y\cdot \na\Phi,\curl \bA\rangle=-3\langle\Phi, \curl \bA\rangle-\langle \Phi,(y\cdot\na)\curl \bA\rangle
\ee
and
\be\la{ek2}
-\langle \na\Phi, (y\cdot \curl \bA)\rangle=\langle\Phi, \curl \bA\rangle+\langle\Phi, \sum\limits_{n=3}^3y_n\na(\curl \bA)_n\rangle
\ee
Finally, 
substituting \eqref{ek1}--\eqref{ek2} into \eqref{ek}, we obtain
\be\nonumber
-\int y\we(\na\Phi \we \curl \bA) dy=-2\langle \curl \bA,\Phi\rangle-\langle y\we\Delta \bA,\Phi\rangle=\langle y\we \bA,\rho\rangle
\ee
So, (\ref{ek0}) is proved.

\section{Acknowledgements}

AK thanks Faculty of Mathematics of Vienna University for the hospitality during the writing the article.
EK thanks  Faculty of Mathematics of Vienna University and Austrian Science Fund FWF for the support
under the project P34177.

\appendix

\protect\renewcommand{\thetheorem}{\Alph{section}.\arabic{theorem}}

\setcounter{equation}{0}
\section{Invariance of the Hamiltonian}\la{aIH}

Here we check (\ref{TRH}).
By (\ref{Hamco}) and definition (\ref{TR}),
\beqn\la{HamcoR}
&&\bH(T(R)\bY)=\fr12\int[\bPi^2+(\curl \bA)^2]dx
\nonumber\\
\nonumber\\
&+&\fr1{2m} [
RP+\langle R\bPi R^{-1}, \na_* R\bA R^{-1}\rangle
-R\blangle \bA(y),\rho(y)\brangle]^2
+\fr1{2I}[R\pi-\blangle y\we R\bA(R^{-1}y),\rho(y)\brangle]^2\qquad\qquad
\eeqn
Using (\ref{rosym}), we obtain that
\be\la{note1}
R\pi-\blangle y\we R\bA(R^{-1}y),\rho(y)\brangle=
R\pi-R\blangle R^{-1}y\we \bA(R^{-1}y),\rho(R^{-1}y)\brangle
=
R[\pi-\blangle z\we \bA(z),\rho(z)\brangle]
\ee
It remains to check that 
\be\la{note2}
\langle R\bPi R^{-1}, \na_* R\bA R^{-1}\rangle=R\langle \bPi , \na_* \bA \rangle.
\ee
Using the definition (\ref{na*}) and (\ref{rosym}), we obtain that for $\xi\in\R^3$
\beqn\la{note3}
&&\langle R\bPi R^{-1}, \na_* R\bA R^{-1}\rangle\cdot\xi:=
\langle R\bPi (R^{-1}y),\xi\cdot \na_y R\bA (R^{-1}y)\rangle
=
\langle R\bPi (R^{-1}y), R[\xi\cdot \na_y]\bA (R^{-1}y)\rangle
\nonumber\\
\nonumber\\
&=&
\langle \bPi (R^{-1}y), [\xi\cdot \na_y]\bA (R^{-1}y)\rangle=
\langle \bPi (R^{-1}y), \fr d{d\ve}\Big|_{\ve=0}\bA (R^{-1}(y+\ve\xi))\rangle
\nonumber\\
\nonumber\\
&=&
\langle \bPi (z), \fr d{d\ve}\Big|_{\ve=0}\bA (z+R^{-1}\ve\xi)\rangle
=
\langle \bPi (z), [R^{-1}\xi\cdot \na_z]\bA (z)\rangle
=
\langle \bPi (z), \na_*\bA (z)\rangle\cdot R^{-1}\xi,
\eeqn
which implies (\ref{note2}).
Finally, (\ref{HamcoR})  and (\ref{note1}), (\ref{note2}) yield
(\ref{TRH}).

\end{document}